\documentclass[floatfix,aps,prl,showpacs,amsmath,nofootinbib,
preprintnumbers,twocolumn]{revtex4}

\usepackage{graphicx}

\begin{document}

\preprint{FERMILAB-PUB-05-335-A}

\title{Post-inflation increase of the cosmological tensor-to-scalar 
perturbation ratio}

\author{N.\ Bartolo}\email{nbartolo@ictp.trieste.it}
   \affiliation{The Abdus Salam International Centre for Theoretical
                Physics, Strada Costiera 11, 34100 Trieste, Italy}

\author{Edward W.\ Kolb}\email{rocky@fnal.gov}
   \affiliation{Particle Astrophysics Center, Fermi
                National Accelerator Laboratory, Batavia, Illinois  60510-0500,
                USA \\ and Department of Astronomy and Astrophysics, Enrico 
		Fermi Institute, University of Chicago, Chicago, Illinois 
		60637-1433, USA}

\author{A.\ Riotto}\email{antonio.riotto@pd.infn.it}
   \affiliation{Department of Physics and INFN
                Sezione di Padova, via Marzolo 8, I-35131 Padova, Italy}

\date{\today}

\begin{abstract}  
We investigate  the possibility that  the amplitude of scalar density
perturbations may be damped after inflation. This would imply that CMB
anisotropies do not uniquely fix the amplitude of the perturbations generated
during inflation and that the present tensor-to-scalar ratio might be larger
than produced in inflation, increasing the prospects of detection of primordial
gravitational radiation. It turns out, however, that the damping of density
perturbations is hard to achieve.
\end{abstract}


\maketitle


Inflation \cite{lrreview} has become the dominant paradigm for
understanding the initial conditions for structure formation and for
Cosmic Microwave Background (CMB) anisotropies. In the inflationary
picture, primordial density and gravitational-wave fluctuations are created
from quantum fluctuations, ``redshifted'' beyond the horizon during an
early period of superluminal expansion of the universe, then
``frozen'' \cite{muk81,bardeen83}. Perturbations at the surface of
last scattering are observable as temperature anisotropies in the CMB,
as first detected by the Cosmic Background Explorer 
satellite \cite{bennett96,gorski96}. The last and most impressive
confirmation of the inflationary paradigm has been recently provided
by data from the Wilkinson Microwave Anisotropy Probe (WMAP)
satellite, which marks the beginning of the precision era of CMB
measurements in space \cite{wmap1}. The WMAP collaboration has
produced a full-sky map of the angular variations of the CMB to
unprecedented accuracy. WMAP data support the inflationary mechanism
as the mechanism for the generation of super-horizon curvature
fluctuations.

The amplitude of cosmological perturbations on large scales can be
expressed in terms of a gauge-invariant variable, $\zeta$,
describing density perturbation on uniform-curvature slices
\cite{bardeen83}: $\zeta=-\psi-H \delta\rho/\dot\rho$,
where the density perturbation $\delta\rho$ and the curvature
perturbation $\psi$ are defined on a generic slicing, $H=\dot a/a$ is
the Hubble rate ($a$ is the scale factor), and the dot stands for
differentiation with respect to cosmic time, $t$.  If the pressure
perturbation is non-adiabatic, the comoving curvature perturbation
$\zeta$ may evolve on super-horizon scales as \cite{bardeen}
\begin{equation}
\label{nad}
\dot\zeta=-\frac{H}{\rho+P}\,\delta P_{\rm nad} ,
\end{equation}
where $P$ is the pressure of the fluid and $\delta P_{\rm
nad}=\delta P- (\dot P/\dot\rho)\delta\rho$ is the non-adiabatic
pressure.

In the standard slow-roll scenario associated with single-field models
of inflation, the observed density perturbations are due to
fluctuations of the inflaton field as it slowly rolls along its
potential $V(\phi)$.  When inflation ends, the inflaton $\phi$
oscillates about the minimum of its potential and decays, thereby
reheating the universe.  As a result of the fluctuations, each region
of the universe goes through the same history, but at slightly
different times. The final temperature anisotropies are caused by the
fact that the duration of inflation was different in different
regions of the universe, leading to adiabatic perturbations.

It is usually assumed that $\zeta$ remains constant on super-horizon
scales, so the curvature perturbation $\zeta$ calculated shortly after
Hubble exit ($k=aH$) during inflation determines the curvature
perturbation until that scale re-enters the Hubble radius during the
subsequent radiation- or matter-dominated phase. Therefore, in the
conventional inflationary scenario, the amplitude of the comoving
curvature perturbation generated during inflation may be directly
related to the large-scale temperature anisotropies through the
Sachs-Wolfe relation $\delta T/T=-(1/5)\,\zeta$.  The power spectrum
of curvature perturbations in slow-roll inflationary models is
predicted to be
\cite{lrreview}
\begin{equation}
{\cal P}_{\zeta}(k)=\frac{k^3}{2\pi^2}\left|\zeta_{\bf k}\right|^2=
\frac{1}{2 M_p^2\epsilon}\left(\frac{H_*}{2\pi}\right)^2
\left(\frac{k}{aH_*}\right)^{n_\zeta-1} ,
\end{equation}
where $n_{\zeta}\simeq 1-2\eta+6\epsilon$ is the scalar spectral index,
$\epsilon= (\dot{\phi}^2/2H^2 M_p^2)\simeq (M_p^2/2)
\left(V'/V\right)^2$ and $\eta=M_p^2\left(V''/V\right)$ are the
slow-roll parameters, $M_p=(8\pi G)^{-1/2}\simeq 2.4\times 10^{18}$
GeV is the reduced Planck mass, and $H_*$ indicates the Hubble rate
during inflation.

The generation of gravitational waves (tensor perturbations) is
another generic prediction of an accelerated de Sitter expansion of
the universe. Gravitational waves, whose possible observation might
come from the detection of the $B$-mode of polarization in the CMB
anisotropies \cite{polreview}, may be viewed as ripples of spacetime
around the background metric
$g_{\mu\nu}=dt^2-a^2(t)\left(\delta_{ij}+h_{ij}\right) dx^i dx^j$.
The tensor $h_{ij}$ is traceless and transverse and has two polarizations
($\lambda = +,\times$).  Since gravitational-wave fluctuations are (nearly)
frozen on superhorizon scales, a way of characterizing them is to
compute their spectrum on scales larger than the horizon.

In the conventional slow-roll inflationary models where the
fluctuations of the inflaton field $\phi$ are responsible for the
curvature perturbations, the power spectrum of the gravitational waves
is $ {\cal P}_{T}(k)=(k^3/2\pi^2)\sum_\lambda\left| h_{\bf k}
\right|^2=(8/M_p^2)\left(H_*/2\pi\right)^2
\left(k/aH_* \right)^{n_T}$, with $n_T=-2\epsilon$.
Since the fractional changes of the power spectra with scales are much
smaller than unity, one can safely consider the power spectra as
roughly constant on the scales relevant for the CMB anisotropy and
define a tensor-to-scalar amplitude ratio 
\begin{equation}
\label{q}
r=0.86\,\frac{{\cal P}_{T}}{{\cal P}_{\zeta}}=13.7 \epsilon\simeq -6.8\, 
n_T .
\end{equation}
This expression provides also a consistency relation between $r$ and
the tensor tilt $n_T$. The present WMAP dataset yields the upper bound
$r\lesssim 0.5$ \cite{ex}.  Since the scale of inflation in slow-roll
models is fixed to be $ V_*^{1/4}\simeq 3.7\,r^{1/4}\times
10^{16}\,{\rm GeV}$, in order to match the observed amplitude of CMB
anisotropies one can already infer an upper bound on the energy scale of
inflation of about $4\times 10^{16}\,{\rm GeV}$.  The corresponding
upper bound on the Hubble rate during inflation, $H_*$, is about
$4\times 10^{14}$ GeV. A positive detection of the tensor modes
through the $B$-mode of CMB polarization (once foregrounds due to
gravitational lensing from local sources have been properly treated)
requires $r \gtrsim 10^{-3}$, corresponding to $V_*^{1/4} \gtrsim
3\times 10^{15}$ GeV and $H_* \gtrsim 2\times 10^{12}$ GeV
\cite{gravex}.

Let us now consider inflation-model phenomenology in the light of
these considerations \cite{lrreview}. Large-field models of inflation
characterized by power-law potentials $V\propto \phi^p$ give a value
of $r=7(1-n_\zeta)p/(p+2)$, making gravitational waves potentially
detectable. However, for small-field models of inflation with
potentials of the type $V\simeq V_0[1-(\phi/M)^2]$, one has $r\simeq
3.5(1-n_\zeta)e^{-(1-n_\zeta)N} \lesssim 1/N$ which is undetectable
($N$ is the number of $e$-folds); for potentials of the form $V\simeq
V_0[1-(\phi/M)^p]$, with $p>2$, one has $r\simeq 7
\,p^2(M/M_p)^{2p/(p-2)}[Np(p-2)]^{-(2p-2)/(p-2)}$, which is 
detectable only if $M$ is much larger than $M_p$. Similar results hold
if $V$ is a mixture of terms, say quadratic at small $\phi$ and
quartic at larger $\phi$, provided that all terms have the same
sign. Hybrid models of inflation, where the inflaton field is
accompanied by another field responsible for most of the potential
energy, also typically have small $r$.

Another way of posing the same problem was suggested in Ref.\
\cite{lythgrav}. The slow-roll paradigm gives, using the definition of
$\epsilon$ and Eq.\ (\ref{q}),
$\left|d\phi/dN\right|/M_p=(r/6.9)^{1/2}$, where $d\phi$ is the change
in the inflaton field in $dN=Hdt\simeq d\,{\rm ln}\,a$ Hubble
times. The range of scales corresponding to the relevant multipoles in
the CMB anisotropy corresponds to $\Delta N\simeq 4.6$, and therefore
the field variation is $(\Delta\phi/M_p)\simeq \left(r/0.2\right)^{1/2}$
\cite{gravex}. This is a minimum estimate because inflation continues for some
number $N$ of $e$-folds of order of 50. The detection of gravitational
waves requires in general variation of the inflaton field of the order
of the Planck scale \cite{lythgrav}, and it is therefore difficult to
construct a satisfactory model of inflation firmly rooted in modern
renormalizable particle theories. There is, therefore, a theoretical
prejudice against the likelihood of observation of gravitational-wave
detection within slow-roll models of inflation where the curvature
perturbation is due to the fluctuations of the inflaton field itself.

This picture sounds fairly pessimistic. The detection of gravitational waves
requires large values of $r$, but the large majority of single-field models of
inflation are characterized by the hierarchy $\epsilon\ll \eta$, and therefore
by tiny values of $r$.  The pessimism relies, however, on the standard
assumption that the comoving curvature perturbation $\zeta$ remains constant on
large scales from the end of the inflationary epoch down to the
matter-dominated era. The goal of this Letter is to raise the possibility that,
within well motivated particle physics scenarios, the amplitude of the
curvature perturbation may be damped during the evolution of the Universe after
inflation. If this is possible, the amplitude of CMB anisotropies on large
scales does not immediately fix the normalization of the comoving curvature
perturbation generated during inflation, $\zeta_{\rm inf}=-
\psi-H\delta\phi/\dot\phi$. In other words, the comoving curvature perturbation
on large scales at the matter-dominated epoch, $\zeta_{\rm m}$, is not equal to
the curvature perturbation during inflation, but it is smaller by a damping
factor $\delta<1$:
\begin{equation}
\zeta_{\rm m}=\delta\,\,\zeta_{\rm inf} .
\end{equation}
What might be the origin of this damping? The total curvature
perturbation is the weighted sum of the individual perturbations,
$\zeta_i=-\psi-H \delta\rho_i/\dot{\rho}_i$, associated with the
various fluids present in the Universe:
\begin{equation}
\zeta=\sum_i\,\frac{\dot{\rho}_i}{\dot \rho}\,\zeta_i\, .
\end{equation}
If after inflation the energy density of the Universe is dominated by
some fluid ``${\rm F}$'' other than radiation, and this fluid is not
perturbed, then $\zeta_{\rm F}=0$ and the total curvature
perturbation is given by $\zeta\simeq
(\dot{\rho}_\gamma/\dot{\rho}_{\rm F})\zeta_\gamma\simeq
({\rho}_\gamma/{\rho}_{\rm F})\zeta_{\rm inf}$, where we have
identified the curvature perturbation of radiation $\zeta_\gamma$ with
the curvature perturbation $\zeta_{\rm inf}$ generated during
inflation. When the fluid ${\rm F}$ decays into relativistic degrees
of freedom commencing conventional radiation domination, the total
curvature perturbation (and therefore the CMB anisotropies) on large
scales is damped with respect to the curvature perturbation generated
during inflation by a factor 
\begin{equation}
\label{delta}
\delta\sim ({\rho}_\gamma/{\rho}_{\rm F})
\end{equation}
computed at the decay time. In simple words, the damping arises
because the fluid releases entropy during its decay, but does not
change the perturbations since $\zeta_{\rm F}=0$.  Therefore, the 
amplitude of CMB anisotropies on large scales does not uniquely fix
the normalization of the comoving curvature perturbation generated
during inflation, and the energy scale of inflation would be fixed
by CMB measurements to be
\begin{equation}
\label{cobenew}
V_*^{1/4}\simeq 7.1\,\frac{\epsilon^{1/4}}{\delta^{1/2}}
\times 10^{16}\,{\rm GeV} .
\end{equation}
This energy can be rather large even in single-field models where the
slow-roll parameter $\epsilon$ is tiny, thanks to the damping factor
$\delta < 1$.

An important fact is that gravitational waves are not similarly damped on
super-horizon scales.  The theoretically predicted value of the
tensor-to-scalar amplitude ratio would then be larger than what is
commonly predicted using Eq.\ (\ref{q}),
\begin{equation}
\label{qq}
r= 0.86\,\frac{{\cal P}_{T}}{{\cal P}_{\zeta}}=13.7  
\frac{\epsilon}{\delta^2}\simeq -6.8 \frac{n_T}{\delta^2} .
\end{equation}
Correspondingly, the consistency relation of Eq.\ (\ref{q}) is
violated and the field variation of the inflaton field is given by
$(\Delta\phi/M_p)\simeq \delta \,\left(r/0.2\right)^{1/2}$.  The
parameter $r$ can be within the detectability range even in those
single-field models of inflation, such as small-field models and
hybrid models, where otherwise the detection would be a hopeless
task. Moreover the detection of gravitational waves would not require
embarrassing variations of the inflaton field as large as the
Planckian scale.

Let us now discuss the possible  source of the damping factor.
Suppose that, beyond the inflaton field, there is another (set of)
complex scalar field(s), generically denoted by $\Phi$. For instance, in the
context of either supergravity or superstring theories, many models of
supersymmetry breaking predict the presence of particles with
Planck-suppressed couplings to ordinary matter \cite{sugra}. These
particles are generically called moduli.  (Warped) Calabi-Yau
compactifications possess a large number of moduli, such as complex
structure moduli, K\"ahler moduli, and dilaton moduli, corresponding
to the deformations of the compact manifold. Flux compactifications
can classically stabilize the moduli \cite{ka,kachru}, and their masses
$m_\Phi$ are typically much larger than the weak scale
\cite{giddings}. This also holds in scenarios such as split
supersymmetry \cite{landscape}, in which the scale of supersymmetry
breaking is orders of magnitude larger than the weak scale and moduli
are expected to acquire masses between $10^{10}$ and $10^{13}$ GeV.

Let us first analyze the case of those moduli which receive a large
mass during inflation because of the supersymmetry breaking 
driven by the  vacuum energy driving inflation \cite{sugra}. 
Barring cancellations, one expects that the mass
of $\Phi$ would be larger than the Hubble rate during inflation. For
example, a supergravity K\"ahler potential of the type $\delta
K=(\Phi^\dagger \Phi\phi^\dagger\phi/M_p^2)$ induces a mass squared for
$\Phi$ equal to $3H_*^2$. For the same reason, a modulus is expected
to be displaced from its present-day minimum. Because of its large
mass, the quantum-mechanical fluctuations of the field $\Phi$ generated
during inflation,
$\zeta_\Phi=-\psi-H \delta\rho_\Phi/\dot{\rho}_\Phi$, can be safely
taken to vanish. After the end of inflation, when the Hubble parameter
becomes smaller than $m_\Phi$, the field $\Phi$ rolls toward its
present value and oscillates about it with an initial amplitude
$\Phi_0$, approximately equal to the distance between the minima of
the effective potential at large $H$ during inflation and small $H$
after inflation. During oscillations the energy density of the
modulus $\Phi$ scales like non-relativistic matter, and is therefore
expected to dominate rather quickly over the radiation energy density
produced during reheating after inflation. At later times the modulus
will decay into relativistic degrees of freedom. 

However, no damping of the comoving curvature perturbation can occur in this
case. The reason is that the modulus starts to oscillate in different parts of
the universe when the local value of the Hubble rate is equal to $m_\Phi$. But
there are local variations in $H$ due to $\zeta_{\rm inf}$, and consequently a
nonvanishing $\zeta_\Phi=\zeta_{\rm inf}$ will be induced. To see it more
concretely, one can analyze the equations for the background modulus field
$\Phi$ and for its fluctuation $\delta\Phi$
\begin{eqnarray}
\label{cc}
\ddot{\Phi}+3H\dot{\Phi}+V'&=&0\, ,\nonumber\\
\ddot{\delta\Phi}+3H\dot{\delta\Phi}+V''\delta\Phi&=&4\dot{\Phi}\dot{\psi}
-2V'\psi\, ,
\end{eqnarray}
where we have assumed that the potential can be expanded linearly,  $\delta
V\sim V'\delta\Phi$. The modulus $\Phi$ may start moving  either when the
universe is still dominated by the energy stored in inflaton oscillations (with
equation of state $w\simeq 0$) or  after reheating when the energy density is
mainly due to   the thermal bath (with equation of state $w\simeq 1/3$).  In
both cases, the gravitational potential $\psi$ can be considered to be
constant. The solution of Eq.\ (\ref{cc}) then reads  $H\delta\Phi\simeq
(2/3(1+w))\dot{\Phi}\psi$. This leads to the initial condition $\zeta_\Phi=
-(5+3w)/(3(1+w))\psi=\zeta_{\rm inf}$ when the modulus starts oscillating, and
no non-adiabatic pressure is available to change the total comoving curvature
perturbation \cite{com}. 

Let us consider, in turn, those moduli which typically do
not receive a large mass during inflation, the stringy axions \cite{sloth}. 
If the K\"ahler
potential does not depend upon the imaginary part of the modulus field, call it
$\sigma$, then the mass of the axion-like field may be  exponentially
suppressed and the condition  $m^2_\sigma\ll H_*^2$ does not require any
particular fine-tuning \cite{bd}. The field $\sigma$ is quantum-mechanically
excited during  inflation, $\delta\sigma\sim (H_*/2\pi)$.  However,  if the
scale $f_\sigma$ appearing in the periodic potential $V(\sigma)$ is smaller
than $H_*$, then the linear  approximation of the fluctuation of the potential
$V(\sigma)$ fails and $\delta\rho_\sigma$ is basically zero. This is because
any long-wavelength fluctuation of the axion field $\sigma$ does not change the
average distribution of $V(\sigma)$ \cite{kl}, thus suppressing $\delta V$. 
This suppression due to the non-linearities of the potential leads to 
\begin{equation}
\delta\rho_{\sigma}\simeq 0
\end{equation}
at the beginning of the radiation phase after inflation. 
The suppression of this entropy perturbations is basically
due to the fact that the perturbations do not follow the background because of
the non-linear potential. However, again, the background
value of the $\sigma$ field will start oscillating around the minimum of the
potential at different times in different separated Hubble volumes
because of the presence of the adiabatic perturbations generated
by the inflaton field. Therefore, adiabatic initial conditions will
be inherited by the field $\sigma$ leading again to 
$\zeta_\sigma=\zeta_\gamma$.

While the cases  analyzed in this short note lead to a negative conclusion,  it
would be most interesting to see if there are (less obvious) sources of   
damping of the primordial curvature perturbation generated during inflation.

We would like to thank M.\ Sloth for very useful correspondence about his
related paper \cite{sloth}  and for readily pointing out an error in the first
version of our draft. We thank A. Linde, D. Lyth, S. Mukhanov 
L. McAllister, M. Sasaki and D. Wands for useful
comments and criticisms. E.W.K.\ is supported in part by NASA grant NAG5-10842
and by the Department of Energy.




\begin{thebibliography}{99}
\bibitem{com} 
Moduli fields may be produced also by thermal scatterings during reheating with
a number density normalized with respect to  the entropy density $s$:
$n_\Phi/s\sim (T_{\rm RH}/M_p) T^3$.  The curvature perturbation of these
thermally produced moduli will be also equal to $\zeta_\gamma=\zeta_{\rm
inf}$.
\end{thebibliography}
\end{document}